\documentclass[a4paper,12pt]{iopart}
\usepackage[UKenglish]{babel} 
\usepackage{iopams}
\usepackage[normalem]{ulem}
\usepackage{graphicx,subfigure}
\usepackage{dcolumn}
\usepackage{hyperref}
\usepackage{xcolor}
\usepackage{amstext}
\usepackage{bbold}
\usepackage{bm}
\usepackage{harvard}

\newcommand{\avg}[1]{\left< #1 \right>} % for average
\newcommand{\ket}[1]{\left| #1 \right>} % for Dirac bras
\newcommand{\bra}[1]{\left< #1 \right|} % for Dirac kets
\newcommand{\braket}[2]{\left< #1 \vphantom{#2} \right|
 \left. #2 \vphantom{#1} \right>} % for Dirac brackets
\renewcommand{\tr}{\text{tr}}

\newcommand{\beeq}{\begin{equation}}
\newcommand{\eneq}{\end{equation}}

\AtEndDocument{\clearpage}

\begin{document}

\paper{Quantacell: Powerful charging of quantum batteries}

\author{Felix C. Binder$^1$, Sai Vinjanampathy$^2$, Kavan Modi$^3$, John Goold$^4$}

\address{$^1$ Clarendon Laboratory, Department of Physics, University of Oxford, Oxford OX1 3PU, United Kingdom} \ead{\mailto{felix.binder@physics.ox.ac.uk}}
\address{$^2$ Center for Quantum Technologies, National University of Singapore, 3 Science Drive 2, 117543 Singapore, Singapore} 
\address{$^3$ School of Physics and Astronomy, Monash University, Victoria 3800, Australia}
\address{$^4$ The Abdus Salam International Centre for Theoretical Physics (ICTP), Trieste, Italy}\ead{\mailto{jgoold@ictp.it}}

\begin{abstract}
We study the problem of charging a quantum battery in finite time. We demonstrate an analytical optimal protocol for the case of a single qubit. Extending this analysis to an array of N qubits, we demonstrate that an N-fold advantage in power per qubit can be achieved when global operations are permitted. The exemplary analytic argument for this quantum advantage in the charging power is backed up by numerical analysis using optimal control techniques. It is demonstrated that the quantum advantage for power holds when, with cyclic operation in mind, initial and final states are required to be separable.
\end{abstract}

%\pacs{xxx}
%\submitto{\NJP}
\maketitle

%*******************************************
%*******************************************
\section{Introduction} 
The maximum work that can be extracted unitarily from a given quantum state $\hat{\rho}$ with respect to a reference Hamiltonian $\hat{H}_0$ was called \emph{ergotropy} by Allahverdyan, Balian, and Nieuwenhuizen \cite{Allahverdyan}. In this extraction scheme the final system state is unique up to degeneracies at the level of the Hamiltonian and belongs to the class of \emph{passive} states~\cite{pusz,lenard}. Such states are characterised by being diagonal in the energy eigenbasis of the Hamiltonian and ordered with decreasing eigenvalues corresponding to increasing energies.

This paradigm of unitary work extraction has recently been extended to scenarios where multi-partite systems are considered. It has been shown by Alicki and Fannes \cite{fannes} that the extractable work can be increased by allowing for entangling operations. Subsequently, Hovhannisyan et al. \cite{Hovhannisyan} demonstrated that, irrespective of an entangling operation being required for optimal work extraction, the entanglement created in the extraction processes can indeed be minimal and even vanish. They argue that the \emph{cost} of this minimisation is a longer process duration and consequently conjecture that there may be a relation between the power of such a process and the entanglement created. This line of reasoning has been extended to explore the work cost in the creation or erasure of more general correlations \cite{giorgi,barcelona2,barcelona3,barcelona4}. In this article we are interested in the inverse process of work extraction: charging a quantum battery. 

We are motivated by the following question: \emph{Are there truly quantum effects or phenomena intrinsic to a specific thermodynamic process that offer an operational advantage over their classical counterpart?} After all thermodynamics should be sensitive to the underlying microscopic description. For example, the efficiency of an engine should always be limited by the Carnot bound irrespective of whether the working medium is comprised of quantum or classical components. However, for non-equilibrium protocols the question is more subtle. For instance very recently it has been demonstrated that it is possible to use non-equilibrium short cuts to adiabaticity to boost the power of engine cycles without compromising efficiency \cite{deng,delcampo}. One may then wonder whether it is in the finite time operation of devices that quantum mechanics offers an advantage. The present article provides an intuitive example of one such process where quantum correlations provide an advantage: When taking the finite time nature of a charging process into account. In this case we demonstrate that the non trivial quantum nature of both the Hamiltonian and the state leads to a substantial advantage. That is, entangling processes, which allow for the creation of quantum correlations, can achieve a higher power than local operations. This is the case even when correlations are only allowed to appear during the process but not in the input and output states.

This paper is structured as follows. In Sec.~\ref{sec:workpower} we outline the general problem of work extraction and powerful driving in a quantum system. We then derive a power-optimal driving for a single qubit -- a case where constraints on the driving are unambiguous -- in Sec.~\ref{sec:qubit}. Sec.~\ref{sec:array} extends the results to an array of $N$ qubits and we demonstrate an advantage in power that stems from the permittance of entangling operations both for an initial pure and an initial thermal state. We conclude in Sec.~\ref{sec:conclusion}.

\section{Charging a quantum battery}
\label{sec:workpower}

A battery is a physical system that stores energy. In this paper we use quantum systems, which store energy in the energy levels and coherences, for this purpose \cite{fannes}. The internal energy of a quantum system (the battery in this case) is then given by $\tr[\hat{\rho} \hat{H}_0]$, where $\hat{\rho}$ is the state of the battery and $\hat{H}_0$ is its internal Hamiltonian. Charging a battery is to change its state from $\hat{\rho}$ to a more energetic state $\hat{\rho}'$ such that $\tr[(\hat{\rho}'- \hat{\rho}) \hat{H}_0] \ge 0$. Conversely, using the battery will take it a lower energy state $\hat{\rho}''$ such that $\tr[(\hat{\rho}'' - \hat{\rho}) \hat{H}_0] \le 0$. We restrict both charging and discharging processes to be cyclic unitary processes. This is achieved by applying an external time-dependent potential $\hat{V}(t)$ for time $T$, in addition to the internal Hamiltonian of the battery $\hat{H}_0$. As a consequence of unitarity the spectrum $\{p_i\}$ of any accessible battery state is fixed \footnote{When comparing two different spectra $\{q_i\}$ and $\{p_i\}$, we note that $\{p_i\}$ allows for both a lower minimum-energy state as also a higher maximum-energy state if it majorises $\{q_i\}$ \cite{majorization}. Consequently, batteries in a pure state have higher capacity than ones in mixed states.}.

The lowest energy state, $\hat{\pi}$, is called a passive state, and analogously the highest energy state, $\hat{\omega}$, a maximally active state; both of them are defined with respect to $\hat{H}_0$. Let us express the internal Hamiltonian with increasing energy levels as 
\begin{equation}
\hat{H}_0:=\sum\epsilon_i\ket{\epsilon_i}\bra{\epsilon_i} \quad \mbox{with}  \quad\epsilon_i \leq \epsilon_{i+1}.
\end{equation}
The passive and active states are respectively
\begin{eqnarray}
&& \hat{\pi}:=\sum_i p_i \ket{\epsilon_i} \bra{\epsilon_i} \quad \mbox{with}  \quad p_i \ge p_{i+1} \quad \mbox{and} \\
&& \hat{\omega}:=\sum_i p_i \ket{\epsilon_i}\bra{\epsilon_i} \quad \mbox{with} \quad p_i \le p_{i+1}.
\end{eqnarray}
Note that all thermal states are passive.

If the battery is in a generic state $\hat{\rho} = \sum_i p_i \ket{r_i}\bra{r_i}$ (with $p_i \geq p_{i+1}$), we can use all available energy and the battery will end up in its passive state. The amount of extractable energy from this battery is called \emph{ergotropy} $\mathcal{W}$. It is defined as the cyclic work extractable with respect to the internal Hamiltonian, $\mathcal{W}:=\tr(\hat{H}_0\hat{\rho})-\tr(\hat{H}_0\hat{\pi})$. Once a quantum state $\hat{\rho}$ has been transformed to a passive state $\hat{\pi}$, no more work is extractable from that state via cyclic unitary transformations. Therefore ergotropy quantifies the available energy in a battery. Conversely, a fully charged battery is in state $\hat{\omega}$ and the system cannot be charged further via a cyclic unitary process. Note that states $\hat{\pi}$, $\hat{\omega}$, and $\hat{\rho}$ are unitarily related to each other. 

In practice it is often desirable to charge a battery quickly, or better: with maximal power. Let us consequently address the question of powerful charging of quantum batteries. For the charging protocol we start with a generic state $\hat{\rho}$  
and apply cyclic, unitary evolution for some optimal time $T$,
\begin{equation}
      \hat{H}_t = \hat{H}_0 + \hat{V}_t  \quad \mbox{with} \quad \hat{V}_t =0\text{, for } t<0\text{ and } t>T,
\end{equation} 
where the index $t$ indicates time-dependence. The protocol duration $T$ is not fixed but rather part of the optimal solution. Here we are not particularly interested in the specific output state $\hat{\rho}'$ but rather in the average work $\avg{W}$ done in the process and, importantly, the average power $\avg{P}={\avg{W}}/{T}$. Alternatively, one could study other process-duration dependent quantities. For instance, maximising $\avg{P} \times \avg{W}$ will guarantee that power is not optimised at the cost of average work. In order to be fully general we thus consider the family of objective functions
\begin{equation}
 \mathcal{F}:=\avg{P}^\alpha\avg{W}^{1-\alpha}=\frac{\avg{W}}{T^\alpha},
 \label{eq:objective}
\end{equation}
where $T$ is the process duration and $0\leq\alpha\leq1$ remains a free parameter. For the extremal case of $\alpha=1$ the expression for power is recovered. Optimisation for $\alpha=0$, on the other hand, recovers the expression for ergotropy as derived in~\cite{Allahverdyan} -- This will be demonstrated in a forthcoming publication~\cite{power} as part of a general optimal control approach to extremising $\mathcal{F}$.

A few remarks are in order:

(1) Cyclicity of the process is an important requirement since we are interested in the power. Consider a non-cyclic process such as a sudden quench, $\hat{H}_{0} \to \hat{H}'$, which does a finite amount of work on the system in an instant and therefore produces infinite power. Such a scenario renders the problem of studying power trivial. If, on the other hand, the process is required to be cyclic then we must have $\hat{H}_{0} \to \hat{H}' \to \hat{H}_0$, and, for a sudden quench, the objective function vanishes both in the numerator and denominator.

(2) One could imagine driving the system infinitely fast between two given states and let the denominator of Eq. \ref{eq:objective} approach zero. Such infinitely powerful driving, however, would be in conflict with so-called quantum speed limits which bound the evolution time by the inverse mean energy~\cite{MargolusLevitin}, or, alternatively, its variance~\cite{MandelstamTamm} and apply to time-dependent unitary evolution between any two quantum states~\cite{Deffner2013}. These speed limits are not prescriptive for determining optimal driving because they depend on the the system state at each instant of time. Alternative bounds, however, have been developed that depend on norms of the driving Hamiltonian~\cite{Raam}.

(3) Such bounds on the Hamiltonian are natural constraints that must be incorporated into the analysis of quantum batteries operating at finite time. In the following sections, we will propose bounds that are external constraints in the sense that they depend on the driving Hamiltonian alone but not the system state. In particular, we will restrict the trace norm of the driving Hamiltonian $\|\hat{H}_t\| \le E_{max}$ with the gauge convention that its lowest eigenvalue is zero. This corresponds to restricting the maximal energy available for external driving. 

%***********************************
%***********************************

\section{Powerful driving for a single qubit}\label{sec:qubit}

Let us now derive an optimal driving with respect to the objective function, Eq.~\ref{eq:objective}, for a single qubit. Without loss of generality we take  $\hat{H}_0=\ket{1}\bra{1}$. We parametrise the driving Hamiltonian $\hat{H}_t$  with control functions $v^x_t, v^y_t, v^z_t$ for the Pauli operators $\hat{\sigma}_x, \hat{\sigma}_y, \hat{\sigma}_z$. The three control functions can be grouped in a vector $\mathbf{v}_t=(v^x_t, v^y_t, v^z_t)^T$ and similarly for the Pauli operators $\boldsymbol{\hat{\sigma}}=(\hat{\sigma}_x, \hat{\sigma}_y, \hat{\sigma}_z)^T$. The instantaneous eigenvalues of $\hat{V}_t$ are $\lambda_t^\pm=\pm|\mathbf{h}_t|=\pm\sqrt{\sum_j(v_t^j)^2}$. A thermodynamically sensible constraint is achieved by bounding the difference between the instantaneous eigenvalues by a maximum value $E_{max}$ -- that is, $\lambda^+_t-\lambda^-_t\leq E_{max}$ for all times $t$ in accordance with the trace norm introduced above.

We now follow recent results, where full analysis of a general time-optimal control of qubit operations was performed \cite{Hegerfeldt,HegerfeldtPRL}. More generally this is related to the so-called \emph{quantum brachistochrone} problem \cite{QuantumBrachistochrone}. These approaches differ from the present analysis in the sense that they are concerned with a process which takes some input state $\hat{\rho}$ to a specific state $\hat{\rho}'$ whereas we are  concerned with the average work that is done in a driving process. Whilst we are interested in preserving the (qubit) state's purity it makes no difference what phase it has. In the Bloch Sphere picture, we are only interested in the state's $z$ coordinate. In this context, $\hat{\rho}$ can be any qubit state (active or passive), i.e., parametrised by angles $\theta$ and $\phi$, and a radius $r$ which remains constant during evolution. The value of $\phi$ has no bearing on the result. As will be seen below this leads to zero driving along $\hat{\sigma}_z$ in the optimal case.

In order to optimise the power we now invoke the von Neumann equation for the state's unitary evolution:
\begin{eqnarray}
  i\frac{d}{dt}\hat{\rho}_t&=[\hat H_0+\hat{V}_t,\hat{\rho}_t]=\frac{1}{2}[\mathbf{v}_t.\boldsymbol{\hat{\sigma}},\mathbb{1}+\mathbf{a}_t.\boldsymbol{\hat{\sigma}}],
\end{eqnarray}
where $\hat{\rho}_t$ is parametrised by its Cartesian decomposition $\mathbf{a}_t$ in terms of Pauli operators, i.e. $\mathbf{a}_t=(a^x_t, a^y_t, a^z_t)^T=r(\sin\theta_t\cos\phi_t, \, \sin\theta_t\sin\phi_t, \, \cos\theta_t)^T$. With Einstein summation convention and using the Levi-Civita symbol the evolution becomes
\beeq
 \frac{d}{dt}{\hat{\rho}_t}=v^j_ta^k_t\epsilon_{jkl}\hat{\sigma}_l.
 \label{eq:rhodotshort}
\eneq
We want to achieve maximum average power $\avg{P}=\avg{W}/T$ (or indeed $\mathcal{F}=\avg{W}/T^\alpha$) over the duration $T$ of the whole process. To reflect cyclicity, work (and consequently: power) is here defined with respect to the initial Hamiltonian $\hat{H}_0$: Charging increases the energy in the state $\hat{\rho}_t$ with respect to the time-independent reference $H_0$. Optimality is hence achieved by first optimising
 $\frac{d}{dt}\tr[\hat{\rho}_tH_0]$
at each instant in time over the driving $\mathbf{v}_t$ and then finding the optimal process duration $T$. Using \autoref{eq:rhodotshort} we obtain
\begin{eqnarray}
   \tr\left(\frac{d}{dt}\hat{\rho}_t\hat{H}_0 \right)&=v^y_ta^x_t-v^x_ta^y_t\\&=(v^y_t\cos\phi_t-v^x_t\sin\phi_t)r\sin\theta_t.
\end{eqnarray}
The optimal protocol under the above constraint is found for
\beeq \label{eq:hxt}\\
 v_t^x=-E_{max}\sin\phi_t, \quad
 v_t^y=E_{max}\cos\phi_t, \quad
 v_t^z=0.
\eneq
This implies that in the optimal case no driving happens in the direction of the reference Hamiltonian. The solution corresponds, unsurprisingly perhaps, to driving along the geodesic with fixed $r$ and $\phi_t=\phi_0$ at constant angular speed $E_{max}$, that is
\begin{equation}
 \theta_t=\theta_0+E_{max}t.
 \label{eq:theta}
\end{equation}
For a fixed input state we now want to optimise the average power, or more generally the function $\mathcal{F}$~\footnote{For a given input state \autoref{eq:theta} already lets us answer two related questions: What is the maximum work that can be achieved in time $T$? What is the minimum time necessary to achieve work $W$ (parametrised by $\theta$)?% This result also resembles the Margolus-Levitin speed limit  \cite{Deffner2013}: $T_{QSL}=\hbar\mathcal{L}(\hat{\rho}',\hat{\rho})/{E_T}$, where $\hat{\rho}$  and $\hat{\rho}'$ are initial and final state respectively. $E_T:=\frac{1}{T}\int_0^T dt |\tr[\hat{\rho} \hat{V}(t)]|$ is the average energy. Note that in the case of pure states the Bures angle $\mathcal{L}(\hat{\rho}',\hat{\rho})$ is indeed given by $\frac{1}{2}|\theta_T-\theta_0|$.
}:

\begin{eqnarray}
    \mathcal{F}&=\frac{\avg{W}}{T^\alpha}=\frac{\tr[\hat{H}_0\hat{\rho}']-\tr[\hat{H}_0\hat{\rho}]}{T^\alpha}\\
	 &=\left(\frac{E_{max}}{\theta_T-\theta_0}\right)^\alpha r\left(\frac{\cos \theta_0}{2}-\frac{\cos \theta_T}{2}\right)\\
	 &=\frac{r}{2T^\alpha}[\cos\theta_0-\cos(\theta_0+E_{max}T)]\label{eq:F}.
\end{eqnarray}
Linearity of the driving time in $\theta_T$ (cf. \autoref{eq:theta}) allows us to choose either variable for optimisation. The optimal solution is found for the driving time $T_m$ that satisfies:
\beeq
  \cos(\theta_0+E_{max}T_m)-\frac{E_{max}T_m}{\alpha}\sin(\theta_0+E_{max}T_m)=\cos\theta_0.
\eneq
% \beeq
%   \frac{E_{max}T_m}{\alpha}=\cot(\theta_0+E_{max}T_m)-\frac{\cos\theta_0}{\sin(\theta_0+E_{max}T_m)}
% \eneq
This equation can be understood geometrically by rewriting it as
\begin{equation}
 \Delta z_T=\frac{E_{max}T}{\alpha}p^{xy}_T,
\end{equation}
which is fulfilled when the qubit has travelled a vertical distance $\Delta z_T$ on the Bloch Sphere which equals the length of its projection onto the $x$-$y$-plane, $p^{xy}_T$, weighed by a factor ${E_{max}T}/{\alpha}$.

\begin{figure}[ht]
\begin{center}
\subfigure[$\;$ Objective function $\mathcal{F}$ for $\alpha=1$ and $\alpha=\frac{1}{2}$]{\label{fig:qubit} \centering\includegraphics[width=0.5\textwidth]{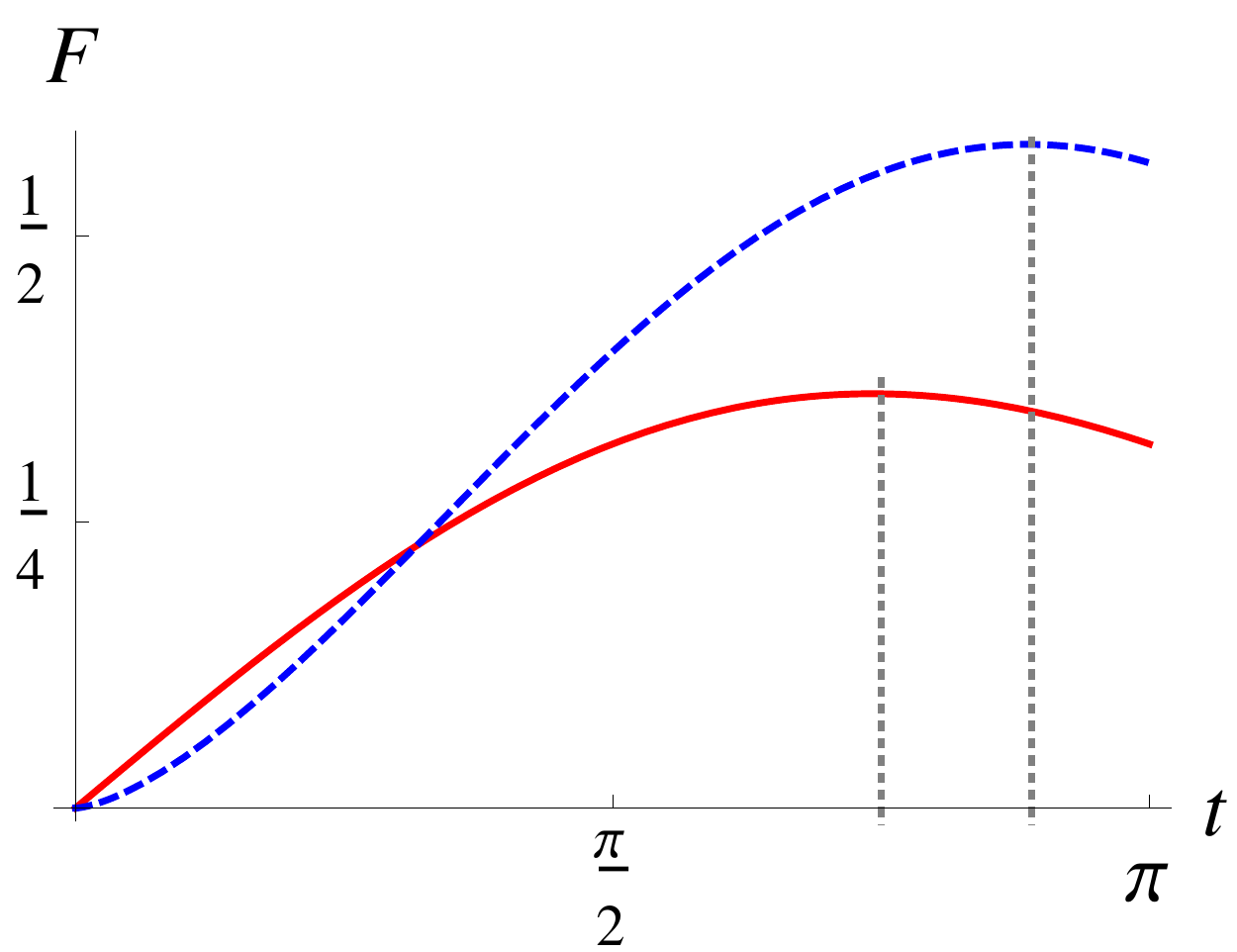}}\hspace{20pt}
\subfigure[$\;$ Path through Bloch Sphere]{\label{fig:maxPpassiveStates}
\centering\includegraphics[width=0.35\textwidth]{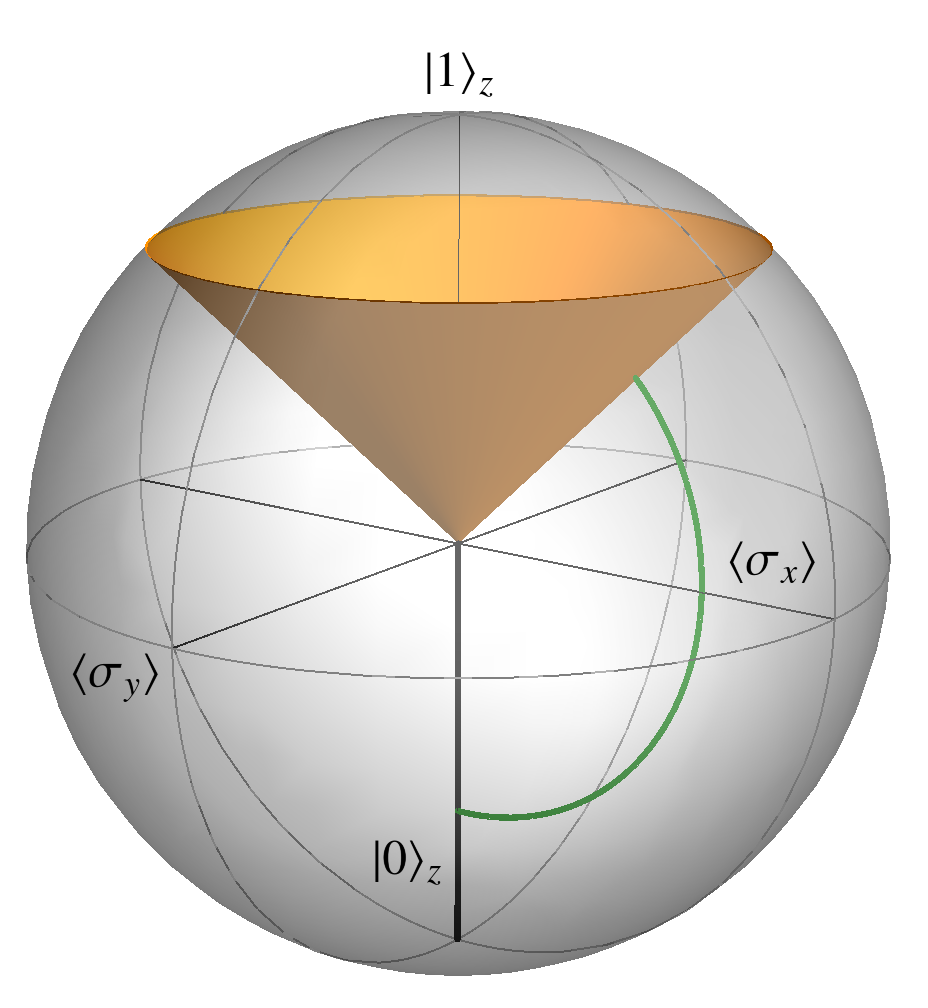}}
\end{center}
\caption{(Colour online.) \textbf{(a)} Taking the generic, thermal input state $\hat{\tau}=\frac{1}{2}[(1\pm r)\ket{0}\bra{0}+(1\mp r)\ket{1}\bra{1}]$ average power $\avg{P}:=\frac{\avg{W}}{T}$ [$\alpha=1$, red, solid], and $\sqrt{\avg{P}\avg{W}}$ [$\alpha=\frac{1}{2}$, blue, dashed] are plotted against $T$. Maximal power is reached for $\theta_f\approx0.74\pi$ in this case, $\avg{P}\avg{W}$ is maximal at $\theta_f\approx0.89\pi$ [grey, dotted lines]. The plot was created with  $E_{max}=1$ for simplicity and the functional values are given in units of $1/r$. \textbf{(b)} For a battery in an initially thermal state (i.e. $\theta_0= \pi$, thick black line), and $E_{max}=1$, this graph shows the cone of states that achieve maximum average power on the Bloch Sphere. Maximal power is reached for $\theta_T \approx 0.74\pi$. Note that a cone only appears for initial states on the $z$-axis. For all other states $\phi$ is fixed and the optimal output state lie on a ray at $\phi_T=\phi_0$. The green line depicts an exemplary trajectory for $r=\frac{2}{3}$ and $\phi=0$.}
\end{figure}

The solution for $\mathcal{F}$ has a non-trivial maximum which is different for different $\alpha$ as can be seen for the example in Fig. \ref{fig:qubit}. 
The cone of optimal output states for initially thermal states is pictured on the Bloch Sphere in Fig. \ref{fig:maxPpassiveStates}. We point out that for certain initial conditions it is possible to achieve finite power, or indeed $\mathcal{F}$, at time $T\to0^+$. This can be regarded as pathological and it is worth pointing out that for $\alpha<1$ and $\theta<\pi$ the optimum is always reached for finite $T$.

%***********************************
%***********************************
\section{Role of entanglement in charging an array of quantum batteries}
\label{sec:array} 
\begin{figure}[ht]
	\centering\includegraphics[width=0.6\textwidth]{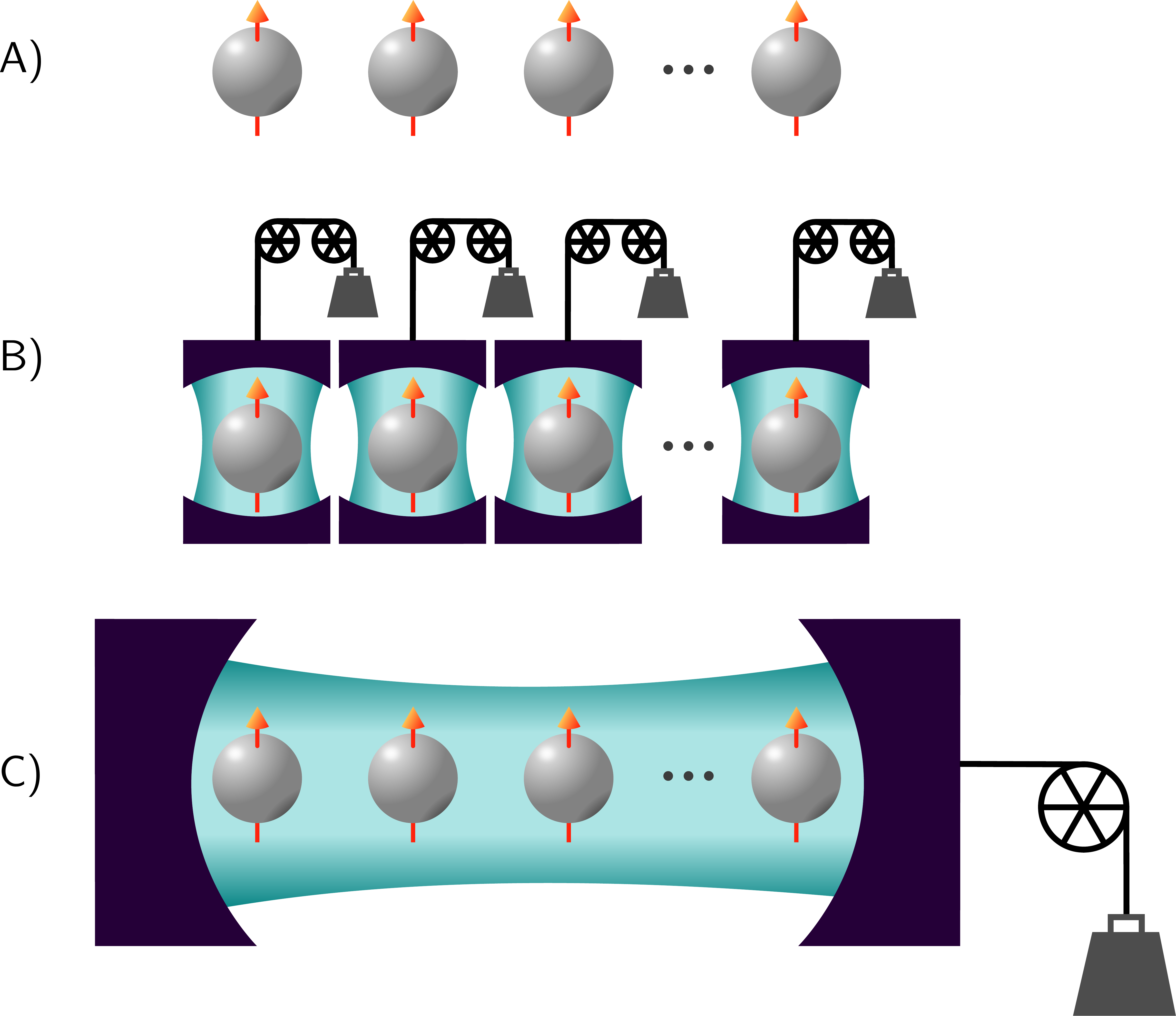}
	 \caption[]{(colour online) Consider an array of qubits (A) that can either be driven in parallel (B), for instance as prescribed by Eq.~\ref{eq:pardriv}, or globally (C), i.e. as given by Eq.~\ref{eq:globdriv}.
	 }
	 \label{fig:array}
\end{figure}

Having found the maximally powerful evolution of a single qubit we now proceed by looking at larger systems, in particular arrays of $N$ qubits, i.e., $\hat{\rho}^{(N)}=\hat{\rho}^{\otimes N}$ as shown in Fig.~\ref{fig:array}. The aim of the following section is to demonstrate that when global, entangling operations are allowed on the array (rather than local, parallel operations for each qubit) an increase in power linear in $N$ can be achieved.

The reference Hamiltonian is the sum of the local Hamiltonians from the single qubit case $\hat{H}_0^{(N)}=\sum_k\ket{1}_{k}\bra{1}_{k}\bigotimes_{j\neq k}\mathbb{1}^{(j)}$ where $k$ labels the $k$th qubit.
For now, we don't want to decrease the purity of the local states (with cyclic operation in mind). We define this -- the condition that there is no reduction in the purity of the marginals at the end of the protocol -- as the condition of \emph{no degradation}. Restricting the analysis to pure states for now, we are interested in going from $\ket{0^{(N)}} := \ket{0}^{\otimes N}$ to $(a\ket{0}+b\ket{1})^{\otimes N}$. Since a full optimisation over all possible unitary transformations is not analytically feasible we will contend ourselves with finding a specific evolution that shows an improvement in power per qubit compared to the single qubit case.

Let us consider full charging such that $\ket{0^{(N)}} \to\ket{1^{(N)}} := \ket{1}^{\otimes N}$. For parallel driving, e.g. Eq.~\ref{eq:hxt}, the time-independent driving Hamiltonian during $0\le t\le T$ is
\begin{equation}
 \hat{H}^{(N)}_{par}=\sum_k^NE_{max}(\cos\phi_0\hat{\sigma}_x^{(k)}-\sin\phi_0\hat{\sigma}_y^{(k)})\bigotimes_{j\neq k}^N\mathbb{1}^{(j)}.
 \label{eq:pardriv}
\end{equation}
It possesses $N+1$ equidistant eigenvalues with the gap between the largest and the lowest eigenvalue given by $N E_{max}$. In order to allow for a fair comparison we will hence equally allow for a maximum energy gap of $E_{max}^{(N)}=N E_{max}$ in the case of global driving. The work done is trivially $\avg{W}=N$  -- that is, the average work per qubit is unity and the driving time is $T_{par}=\pi/E_{max}$. Global driving is achieved by evolution with~\footnote{This global driving is in fact optimal for the current scenario. Generally, global driving between any two states $\ket{a}$ and $\ket{b}$ is optimised by evolution with $\hat H_d\propto\ket{a}\bra{b}+\ket{b}\bra{a}$ where the proportionality is given by the external constraint.}
%\footnote{For two qubits we can express $\hat{\sigma}^{(2)}_x=\frac{1}{2}(\hat{\sigma}_x^{\otimes 2}-\hat{\sigma}_y^{\otimes 2})$.
\begin{equation}
      \hat{H}_{global}= E_{max}^{(N)}\hat{\sigma}^{(N)}_x := E_{max}^{(N)}\left(\ket{1^{(N)}}\bra{0^{(N)}}+\ket{0^{(N)}} \bra{1^{(N)}}\right)
      \label{eq:globdriv}
\end{equation}
for a duration $T=\pi/E_{max}^N$. The average work in this process is again $\avg{W}=N$. The average power consequently reads
\beeq
 \avg{P}=\frac{E^{(N)}_{max}}{\pi}N
 \label{eq:globalP}
\eneq
and depends on the driving constraint $E_{max}^{(N)}$ for $\hat{\sigma}^{(N)}_x$. Permitting for the same constraint as in the parallel case, that is $E_{max}^{(N)}=N E_{max}$, an $N$-fold increase in power per qubit is achieved. Interestingly, the advantage from correlations appears even though initial and final states are separable.

The quantum speedup observed here can be understood in part as a consequence of the shorter distance that has to be travelled through state space when entangling operations are permitted. To calculate and compare the path length in state space between the global and local pure state case (i.e., $\ket{0^{(N)}} \to \ket{1^{(N)}}$) we first note that in the global case the proposed evolution does in fact prescribe a path along a geodesic. As a consequence the path length is directly given by the Bures angle $\mathcal{L}$ between the input and the output state. For pure states the Bures angle is equal to the Fubini-Study distance and we directly see~\cite{Bengtsson}
\begin{equation}
 \mathcal{L} \left(\ket{0^{(N)}}, \ket{1^{(N)}} \right) =\arccos \left|\braket{0^{(N)}}{1^{(N)}} \right|
 =\frac{\pi}{2}
\end{equation}
In the case of parallel driving we note that the square of the line element is additive under tensor product. That is, for $\ket{\psi_{global}}=\otimes_j\ket{\psi_{j}}$, $ds_{global}^2=\sum_jds_j^2$. Accordingly the path length in this case scales with $\sqrt{N}$.

\begin{figure}[ht]
	\centering\includegraphics[width=0.8\textwidth]{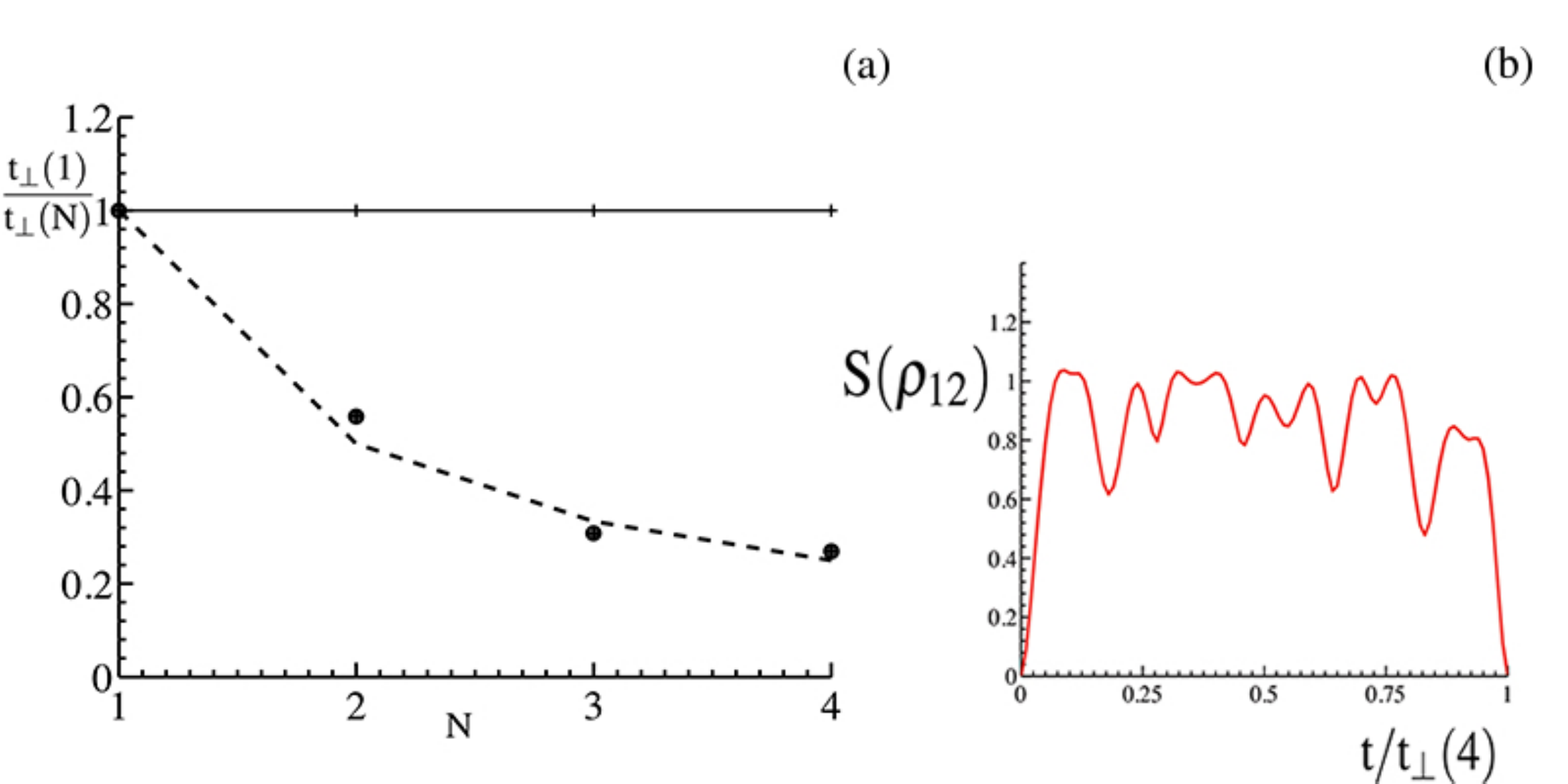}
	 \caption[]{(colour online) Numerical optimization results for minimum time $t_\perp$ for complete charging of $N$ qubits, where the state of the battery rotates from  $\ket{0^{(N)}}$ to $\ket{1^{(N)}}$. The maximum eigenvalue of the Hamiltonian is scales linearly with the number of qubits, and we note that the full charging time $t_\perp$ is commensurate with $1/N$. The solid line represents using the single qubit charging protocol for all $N$ qubits. The dashed line represents the $1/N$ line, whereas the circles represent the solution to numerical optimisation. The entanglement dynamics for the four qubit numerical optimization is shown in (b), by plotting the entropy of the reduced state obtained by tracing over the last two qubits. We note that at the beginning and at the end of the dynamics, the qubits are disentangled, indicated by vanishing entropies. For the numerical optimisation the eigenvalues of the instantaneous Hamiltonian were bounded between $0$ and $E_max=1$. Since the absolute scaling $E_max$ of the Hamiltonian is arbitrary, the single qubit time serves as a  normalization to see the behaviour exhibited.}
	 \label{fig:T_vs_N_Optim}
\end{figure}

For comparison, let us numerically study the same charging process, i.e., namely going from $\ket{0^{(N)}} \to\ket{1^{(N)}}$. Fig.~\ref{fig:T_vs_N_Optim} shows results of a numerical optimisation of full charging between the number of qubits in (a), plotted on the $x$-axis, and the time that the quantum system takes to rotate between the states. The minimum time represents an optimisation over all possible time-independent Hamiltonians such that their eigenvalues are bounded by the same constraint as above (a factor linear in $N$). We see that the charging time is commensurate with $1/N$ behaviour. We note that the dots represent the numerical optimisation of full charging, whereas the dashed line represents the $1/N$ behaviour, as a guide to the eye. The solid line represents parallel charging of all $N$ qubits using the single qubit protocol. Finally, the plot in (b)  represents the evolution of the entanglement of the quantum system for the four qubit numerical solution. Plotted is the reduced entropy of the first two qubits, showing the generation of entanglement across the 12--34 bipartition. Note that as the system approaches the final state, corresponding to the optimal value of power, the qubits disentangle. This is commensurate with the demand that all batteries be available for individual use at the final time with loss of purity. This exemplifies the role of entanglement during the protocol as being responsible for the quantum speedup.

\section{Conclusion}
\label{sec:conclusion}
In this article we have given a full derivation of a qubit protocol that achieves maximum power when charging a quantum battery -- a work qubit -- under constrained driving. With cyclic operation in mind we allowed for unitary evolution, hence keeping the state's purity constant. In extension we then examined the charging of an array of $N$ work qubits. Using a specific exemplary evolution we demonstrated $N$-fold advantage in power per work qubit. This example was presented under the operational constraint that the purity of the state must again be conserved. This highlights the potential for quantum enhancement of devices working under non equilibrium conditions. For future work it will be interesting to extend this scenario and allow for \emph{degradation} -- that is, an extra purifying stage is added after the discharging of the battery array to make up for any loss in purity during the global charging process. Furthermore, in order to allow for realistic noise and decoherence processes it will be of paramount importance to generalise the approach to open systems \cite{binder}. We hope that this work will inspire further investigations into the advantages of quantum effects in finite-time thermodynamic processes. 

%\begin{acknowledgments}
\ack
The authors thank R. Fazio, V. Vedral, M. Taddei, L. Davidovich, and M. Perarnau-Llobet for insightful comments. FB acknowledges funding by the Rhodes Trust. Centre for Quantum Technologies is a Research Centre of Excellence funded by the Ministry of Education and the National Research Foundation of Singapore. This work was partially supported by the COST Action MP1209. The authors hope that this article will inspire Tesla Motors to start developing quantum batteries for their vehicles.
%\end{acknowledgments}

% *******************************************
\section*{References}
\bibliography{NJP_references}
\bibliographystyle{jphysicsB}
\end{document}